\RequirePackage{lineno}
\documentclass[prd,aps,noeprint,superscriptaddress,twocolumn,amsmath,amssymb]{revtex4-2}
\usepackage[colorlinks,linkcolor=blue,anchorcolor=blue,citecolor=blue,urlcolor=blue]{hyperref}
\usepackage{lineno}
\usepackage{graphicx}
\usepackage{subfigure}
\usepackage{bm}
\usepackage{overpic}
\usepackage{verbatim}
\usepackage{rotating}
\usepackage{tabularx}
\usepackage{array}
\usepackage{setspace}
\usepackage{dcolumn}
\newcolumntype{C}[1]{>{\centering}p{#1}}
\newcolumntype{L}[1]{>{\raggedright}p{#1}}
\newcolumntype{R}[1]{>{\raggedleft}p{#1}}
\newcolumntype{P}{@{$~\pm~$}}
\newcommand{\unitegev}{~{\rm GeV}}
\newcommand{\unitemev}{~{\rm MeV}}
\newcommand{\unitpgev}{~{\rm GeV}/c}

\newcommand{\unitmmev}{~{\rm MeV}/c^2}

\newcommand{\ra}{\rightarrow}
\newcommand{\ee}{e^+e^-}
\newcommand{\pip}{\pi^+}
\newcommand{\pim}{\pi^-}
\newcommand{\piz}{\pi^0}

\newcommand{\chisq}{\chi^{2}}
\newcommand{\chisqfour}{\chi^{2}_{\rm 4C}}

\newcommand{\born}{\sigma^{\rm B}}

\newcommand{\luminosity}{647~{\rm pb}^{-1}}
\newcommand{\energy}[1][2.1250]{\sqrt{s}=#1\unitegev}

\newcommand{\ppp}{\pip\pim\piz}
\newcommand{\pipipi}{\ppp}
\newcommand{\process}[1]{\ee\ra#1}

\newcommand{\processppp}{\process{\ppp}}

\newcommand{\rhopi}{\rho\pi}
\newcommand{\rhostarpi}{\rho(1450)\pi}
\newcommand{\processrho}{\ee\ra\rhopi}
\newcommand{\processrhostar}{\ee\ra\rhostarpi\ra\ppp}

\def\babar{\mbox{\slshape B\kern-0.1em{\small A}\kern-0.1em B\kern-0.1em{\small A\kern-0.2em R}}}

\newcommand{\resultpppmass}{2119}
\newcommand{\resultpppmasserrsta}{11}
\newcommand{\resultpppmasserrsys}{15}
\newcommand{\resultpppwidth}{69}
\newcommand{\resultpppwidtherrsta}{30}
\newcommand{\resultpppwidtherrsys}{5}
\newcommand{\resultpppsignificance}{5.9}
\newcommand{\resultpppcodemass}{$M=\resultpppmass\pm\resultpppmasserrsta\pm\resultpppmasserrsys\unitmmev$}
\newcommand{\resultpppcodewidth}{$\varGamma=\resultpppwidth\pm\resultpppwidtherrsta\pm\resultpppwidtherrsys\unitemev$}
\newcommand{\resultpppcodesignificance}{$\resultpppsignificance\sigma$}
 
\begin{document}
% \linenumbers
\normalsize
\parskip=5pt plus 1pt minus 1pt
\title{\boldmath Study of $\processppp$ at $\sqrt{s}$ from 2.00 to 3.08~GeV at BESIII}
\author{
    M.~Ablikim$^{1}$, M.~N.~Achasov$^{4,b}$, P.~Adlarson$^{75}$, X.~C.~Ai$^{81}$, R.~Aliberti$^{35}$, A.~Amoroso$^{74A,74C}$, M.~R.~An$^{39}$, Q.~An$^{71,58}$, Y.~Bai$^{57}$, O.~Bakina$^{36}$
% ...
    \\
    \vspace{0.2cm}
    (BESIII Collaboration)\\
    \vspace{0.2cm} {\it
        $^{1}$ Institute of High Energy Physics, Beijing 100049, People's Republic of China\\
% ...
$^{4}$ Budker Institute of Nuclear Physics SB RAS (BINP), Novosibirsk 630090, Russia\\
% ...
$^{35}$ Johannes Gutenberg University of Mainz, Johann-Joachim-Becher-Weg 45, D-55099 Mainz, Germany\\
        $^{36}$ Joint Institute for Nuclear Research, 141980 Dubna, Moscow region, Russia\\
% ...
$^{39}$ Liaoning Normal University, Dalian 116029, People's Republic of China\\
% ...
$^{57}$ Southeast University, Nanjing 211100, People's Republic of China\\
        $^{58}$ State Key Laboratory of Particle Detection and Electronics, Beijing 100049, Hefei 230026, People's Republic of China\\
% ...
$^{71}$ University of Science and Technology of China, Hefei 230026, People's Republic of China\\
% ...
$^{74}$ University of Turin and INFN, (A)University of Turin, I-10125, Turin, Italy; (B)University of Eastern Piedmont, I-15121, Alessandria, Italy; (C)INFN, I-10125, Turin, Italy\\
        $^{75}$ Uppsala University, Box 516, SE-75120 Uppsala, Sweden\\
% ...
$^{81}$ Zhengzhou University, Zhengzhou 450001, People's Republic of China\\
        \vspace{0.2cm}
% ...
$^{b}$ Also at the Novosibirsk State University, Novosibirsk, 630090, Russia\\
% ...
    }
}
\date{\today}
\begin{abstract}
    {
        With the data samples taken at center-of-mass energies from 2.00 to 3.08 GeV with the BESIII detector at the BEPCII collider, a partial wave analysis on the $\processppp$ process is performed. The Born cross sections for $\processppp$ and its intermediate processes $\ee\ra\rho\pi$ and $\rho(1450)\pi$ are measured as functions of $\sqrt{s}$. The results for $\processppp$ are consistent with previous results measured with the initial state radiation method within one standard deviation, and improve the uncertainty by a factor of ten. By fitting the line shapes of the Born cross sections for the $\ee\ra\rho\pi$ and $\ee\ra\rho(1450)\pi$, a structure with mass \resultpppcodemass\ and width \resultpppcodewidth\ is observed with a significance of \resultpppcodesignificance, where the first uncertainties are statistical and the second ones are systematic. This structure can be intepreteted as  an excited $\omega$ state.
    }
\end{abstract}
\maketitle
\section{Introduction}
{
The anomalous magnetic moment of the muon $a_{\mu}\equiv~(g-2)_{\mu}/2$ plays a very important role to test with high precision the Standard Model (SM) calculations and search for new physics signatures~\cite{3_2_g2_white_paper}.
Recent results lead to a factor of two improvement in precision of the experimental world average~\cite{3_2_g2_2023}, which brings to a reevaluation of the 4.2 standard deviation from the SM predictions obtained by combining results of the E989 and E821 Collaborations~\cite{3_2_g2_2006, 3_2_g2_2021}.
%Recently, the new experimental world average, which represents a factor of two improvement in precision~\cite{3_2_g2_2023}, changes the situation that the combined result of the E989 and E821 Collaborations is greater than the SM prediction by 4.2 standard deviations~\cite{3_2_g2_2006, 3_2_g2_2021}.
The lattice QCD calculations~\cite{3_2_g2_3} reduce the discrepancy with the experimental measurement of the hadron vacuum polarization (HVP) term $a_{\mu}^{\rm HVP}$. Further studies of the processes included in the HVP contribute to reduce the $a_{\mu}^{\rm HVP}$ uncertainty. Among them, the contribution of the \mbox{$\processppp$} process at energies below 2~GeV is second only to that of the $\ee\ra\pip\pim$ channel. The widely used generator for this process is Phokhara~\cite{4_2_gen_phokhara}, based on $\babar$'s results~\cite{3_2_babar_1, 3_2_babar_2}. A more precise study of the process \mbox{$\processppp$}, based on amplitude analysis, is needed to improve Phokhara thus the measurements of $a_{\mu}^{\rm HVP}$ below 2~GeV.

Previously, the process $\ee\ra\pipipi$ was studied at center-of-mass energies ($\sqrt{s}$) from 0.6 to 2.0~GeV in direct $\ee$ annihilations at SND~\cite{3_2_snd1,3_2_snd2,3_2_snd3,3_2_snd4,3_2_snd5} and CMD-2~\cite{3_2_cmd1,3_2_cmd2}, and from threshold to 3.0~GeV by \babar~\cite{3_2_babar_1, 3_2_babar_2} and BESIII~\cite{3_2_bes1} with the initial state radiation (ISR) technique. Among these studies, only the SND experiment performed an amplitude analysis to estimate the $\ee\ra\rhopi$ and $\rhostarpi$ contributions at $\sqrt{s}$ from 1.15 to 2.0~GeV~\cite{3_2_snd1,3_2_snd2,3_2_snd3,3_2_snd4,3_2_snd5}. Such studies are now extended within this work in the energy region from 2.0 to 3.08~GeV with data collected by BESIII. Throughout this paper, $\rho$ denotes $\rho(770)$ and $\rhopi$ incorporates $\rho^{0}\piz$, $\rho^{+}\pim$ and $\rho^{-}\pip$ with fractions that satisfy isospin conservation.

The measurement of Born cross sections for the \mbox{$\processppp$} process is also critical to search for potential excited vector mesons. In $\ee$ collisions at energies below 2 GeV, the vector mesons $\rho$, $\omega$, and $\phi$, as well as their excited states, are produced abundantly. However, above $\energy[2.0]$, some higher lying excited states are not fully identified, and additional experimental measurements are needed to reveal the natures of potential excited states such as $\omega(2205)$~\cite{2_omega_2205}, $\omega(2290)$~\cite{2_omega_2290}, and $\omega(2330)$~\cite{2_omega_2330}, since their measured widths are significantly different from theoretical predictions. In the meantime, recent cross section measurements of \mbox{$\ee\ra\omega\pip\pim$}~\cite{2_bes_opm} and \mbox{$\ee\ra\omega\piz\piz$} processes at BESIII~\cite{2_bes_opp} and \mbox{$\ee\ra\omega\pi\pi$} at \babar~\cite{3_1_babar_opm, 3_1_babar_opp} did not improve the situation due to the large uncertainty from continuum backgrounds. The structure observed with significance larger than $5\sigma$ in the \mbox{$\ee\ra\omega\pi\pi$} cross section line shape was found to be less significant in the line shape of intermediate processes, which were extracted from an amplitude analysis quoted as partial wave analysis (PWA)~\cite{5_1_pwa_theory}.

With the data samples collected with the BESIII detector at nineteen center-of-mass energies from 2.00 to 3.08~GeV, corresponding to a total integrated luminosity of $\luminosity$, the Born cross sections of exclusive hadron channels have been measured and excited meson states have been searched for with the processes $\ee\ra\eta\prime(958)\pip\pim$~\cite{3_3_etappipi}, $K^{+}K^{-}$~\cite{3_3_kk}, $\omega\piz$~\cite{3_3_omegaetaomegapi}, $\omega\eta$~\cite{3_3_omegaetaomegapi}, $K^{+}K^{-}\piz\piz$~\cite{3_3_kkpipi}, $\phi\pi^{+}\pi^{-}$~\cite{3_1_phipipi_bes}, $K_{S}^{0}K_{L}^{0}$~\cite{3_3_kskl}, $\phi\eta$~\cite{3_3_phieta}, $\phi\eta\prime(958)$~\cite{3_3_phietap}, and $\phi K^{+}K^{-}$~\cite{3_3_phikk}. In this paper, the Born cross sections of the process \mbox{$\processppp$} and its intermediate processes (through $\rhopi$ and $\rhostarpi$) are measured with the same data samples. The dynamic of the process \mbox{$\processppp$} are studied in this energy region to improve the Phokhara generator.

\par
}

\section{Detector and data sample}
{
The BESIII detector~\cite{4_1_detector_1} records symmetric $e^+e^-$ collisions provided by the BEPCII storage ring~\cite{4_1_detector_2}, which operates in the center-of-mass energy range from 2.0 to 4.95 GeV. The cylindrical core of the BESIII detector covers 93\% of the full solid angle and consists of a helium-based multilayer drift chamber~(MDC), a plastic scintillator time-of-flight system~(TOF), and a CsI(Tl) electromagnetic calorimeter~(EMC), which are all enclosed in a superconducting solenoidal magnet providing a 1.0~T magnetic field. The solenoid is supported by an octagonal flux-return yoke with resistive plate counter muon identification modules interleaved with steel. The charged-particle momentum resolution at $1~{\rm GeV}/c$ is $0.5\%$, and the specific ionization energy loss ${\rm d}E/{\rm d}x$ resolution is $6\%$ for electrons from Bhabha scattering. The EMC measures photon energies with a resolution of $2.5\%$ ($5\%$) at $1$~GeV in the barrel (end cap) region. The time resolution in the TOF barrel region is 68~ps, while that in the end cap region is 110~ps.

Simulated data samples produced with a {\sc geant4}-based~\cite{4_2_gen_geant4} Monte Carlo (MC), which includes the geometric description of the BESIII detector and the detector response, are used to determine detection efficiencies and to estimate backgrounds. The known decay modes are modeled with {\sc evtgen}~\cite{4_2_gen_evtgen1, 4_2_gen_evtgen2} using branching fractions taken from the Particle Data Group (PDG)~\cite{1_PDG}. Final state radiation from charged final state particles is incorporated using the {\sc photos} package~\cite{4_2_gen_photons}, and the ISR is incorporated using the ConExc~\cite{4_2_gen_conexc} generator using as input the measured cross sections in an iterative procedure.

The signal MC samples of the \mbox{$\processppp$} process with \mbox{$\piz\ra\gamma\gamma$} are generated using a phase space model and weighted according to the PWA results from the data to estimate the detection efficiency, together with its intermediate processes \mbox{$\processrho$} and $\rhostarpi$. To study the potential background, the inclusive MC samples are generated using a hybrid generator~\cite{4_2_gen_hybrid}, which includes hadronic, Bhabha scattering, di-muon, and di-photon processes.
}

\section{Event selection and background analysis}
{
The candidate events are selected with two reconstructed charged tracks and at least two reconstructed photons.
The charged tracks detected in the MDC are required to satisfy \mbox{$|\rm{cos\theta}|<0.93$}, where $\theta$ is defined as the polar angle with respect to the $z$-axis, which is the symmetry axis of the MDC. The distance of closest approach to the interaction point must be less than 10\,cm along the $z$-axis and less than 1\,cm in the transverse plane.
The $E/p$ ratio of the charged track, defined as the ratio of the energy deposited in the EMC to the momentum measured by the MDC, is required to be less than 0.8 to suppress the background from Bhabha events.
Since the opening angle ($\theta_{+-}$) between the two charged tracks in Bhabha and di-muon events is close to $180^\circ$ in the center-of-mass frame, the condition $\theta_{+-}<160^\circ$ is required to suppress the remaining background from Bhabha and di-muon events.

Photon candidates are identified using showers in the EMC. The deposited energy of each shower must be larger than 25~MeV in the barrel region \mbox{($|\!\cos\theta|<0.80$)} and larger than 50~MeV in the end cap region \mbox{($0.86<|\!\cos \theta|<0.92$)}. To exclude showers that originate from charged tracks, the opening angle between the position of each shower in the EMC and the closest extrapolated charged track must be greater than $10^\circ$. To suppress electronic noise and showers unrelated to the event, the difference between the EMC time and the event start time is required to be within [0, 700]~ns.

To further suppress background, a four-constraint (4C) kinematic fit imposing energy-momentum conservation is performed under the hypothesis $\ee~\ra~\pip\pim\gamma\gamma$, and the condition $\chisqfour<50$ is required. For events with more than two photon candidates, the combination with the smallest $\chisqfour$ is retained.

The two reconstructed photons are required to have a helicity angle ($\theta_{\rm h}$) satisfying $|{\rm cos}\theta_{\rm h}|<0.8$ to suppress the backgrounds associated with remaining photons, defined as the angle between the momentum of the $\piz$ in the laboratory coordinate system and the momentum of the photon in the $\piz$ coordinate system.

The $\gamma\gamma$ invariant mass distribution of \mbox{$\processppp$} candidates is shown in Fig.~\ref{figure:signalyield}; in this distribution, the $\piz$ signal region is selected as \mbox{$|M_{\gamma\gamma}-M^{\rm PDG}_{\pi^0}|<15\unitmmev$}, corresponding to about \mbox{$\pm3\sigma$} around the known $\piz$ mass~\cite{1_PDG}. The selected candidate events are used to perform the PWA, while the events in the sideband region \mbox{$|M_{\gamma\gamma}-M^{\rm PDG}_{\pi^0}|\in[30, 60]\unitmmev$} are used to estimate the background contribution. No peaking background is observed in the $\piz$ signal region. The background levels of data used in PWA are estimated by sidebands to be $\sim1\%$ across different energy points. The number of events in the $\piz$ signal region and in the sideband region, together with the estimated background level, are listed in Table~\ref{table:background_level}.

\begin{figure}[hbpt]
    \begin{center}
        \includegraphics[width=0.48\textwidth]{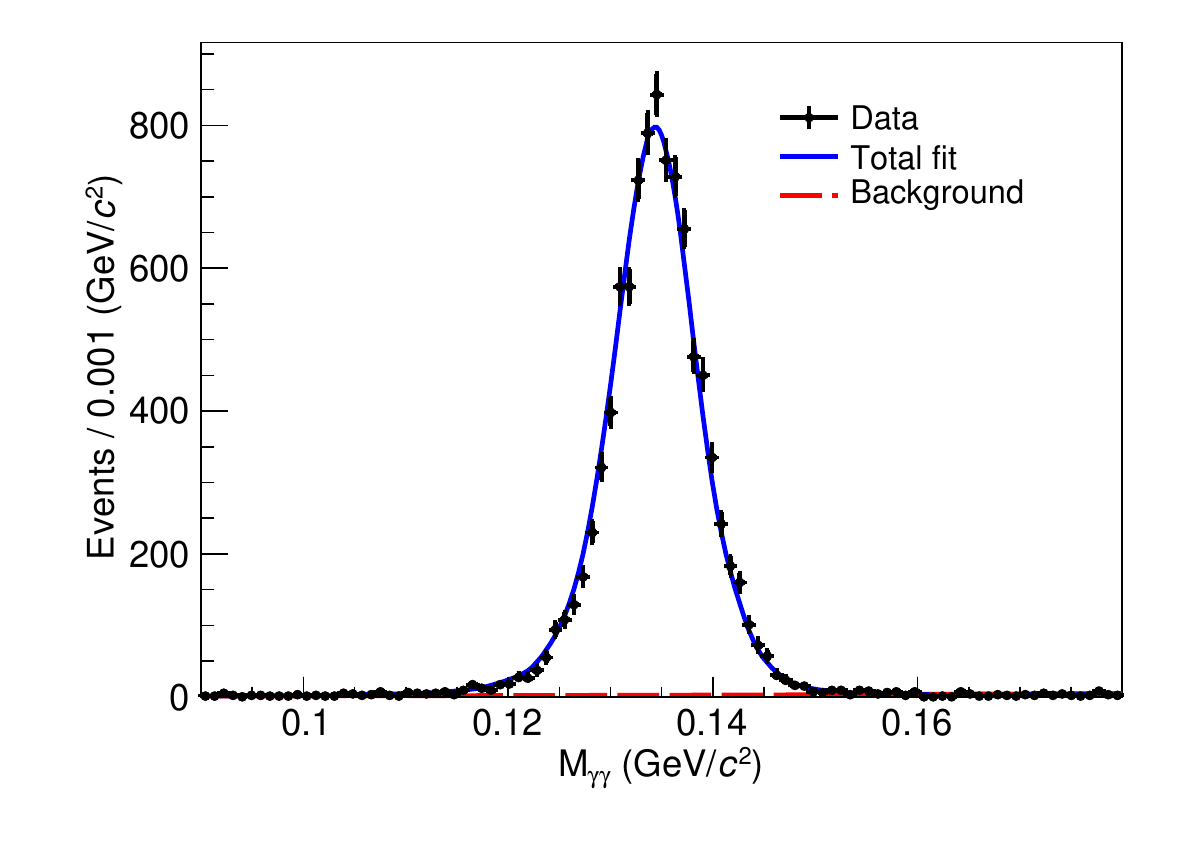}
        \caption{The $\gamma\gamma$ invariant mass distribution at \mbox{$\energy$}. The black dots with error bars represent data. The blue solid line represents the total fit function. The red dashed line represents the background.}
        \label{figure:signalyield}
    \end{center}
\end{figure}

\begin{table}[!htbp]
    \begin{center}
        \caption{The number of events in signal ($N_{\rm S}$) and sideband ($N_{\rm B}$) region, together with the estimated background level, where the number of background events is normalized by a factor of 2.}
        \label{table:background_level}
        \begin{small}
            \begin{tabular}{c|c|c|c}
                \hline\hline
                $\sqrt{s}\ {\rm GeV}$ & $N_{\rm S}$ & $N_{\rm B}$ & $N_{\rm S}/N_{\rm B} (\%)$ \\
                \hline
                2.0000                & 936         & 10          & 1.07                       \\
                2.0500                & 254         & 2           & 0.79                       \\
                2.1000                & 972         & 8           & 0.82                       \\
                2.1250                & 9408        & 76          & 0.81                       \\
                2.1500                & 239         & 1           & 0.42                       \\
                2.1750                & 868         & 4           & 0.46                       \\
                2.2000                & 1032        & 5           & 0.48                       \\
                2.2324                & 842         & 11          & 1.31                       \\
                2.3094                & 1170        & 12          & 1.03                       \\
                2.3864                & 949         & 10          & 1.05                       \\
                2.3960                & 2824        & 36          & 1.27                       \\
                2.6444                & 768         & 11          & 1.43                       \\
                2.6464                & 753         & 13          & 1.73                       \\
                2.9000                & 1335        & 30          & 2.25                       \\
                2.9500                & 154         & 4           & 2.60                       \\
                2.9810                & 198         & 2           & 1.01                       \\
                3.0000                & 165         & 2           & 1.21                       \\
                3.0200                & 151         & 6           & 3.97                       \\
                3.0800                & 610         & 21          & 3.44                       \\
                \hline\hline
            \end{tabular}
        \end{small}
    \end{center}
\end{table}

Figure~\ref{figure:dalitz} shows the Dalitz plots of \mbox{$\processppp$} candidate events at $\energy$ in data and as predicted by the Phokhara generator. The model implemented in the generator fails clearly to describe the intermediate states produced in the process, thus it cannot be employed in this work to determine the detection efficiency.

\begin{figure}[hbpt]
    \begin{center}
        \begin{overpic}[width=0.23\textwidth]{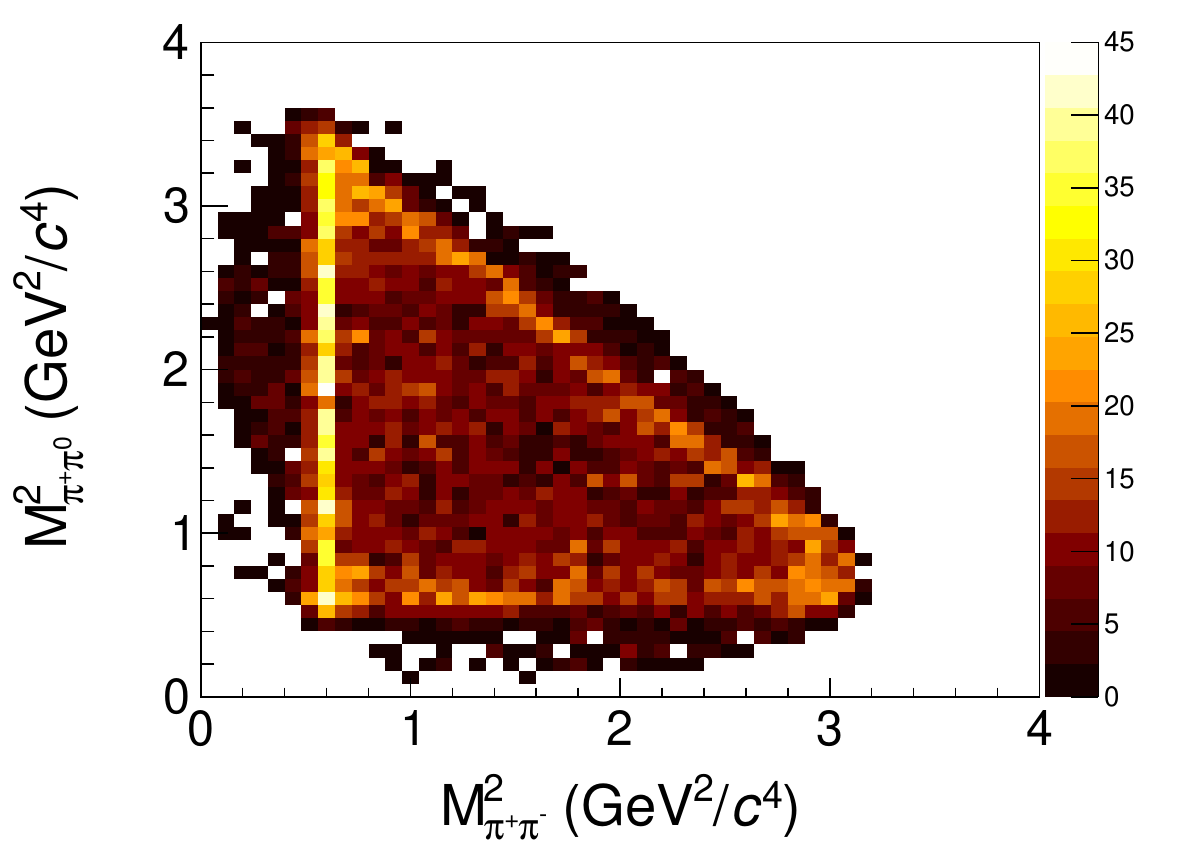}
            \put(60,60){\small\bfseries{(a)}}
        \end{overpic}
        \begin{overpic}[width=0.23\textwidth]{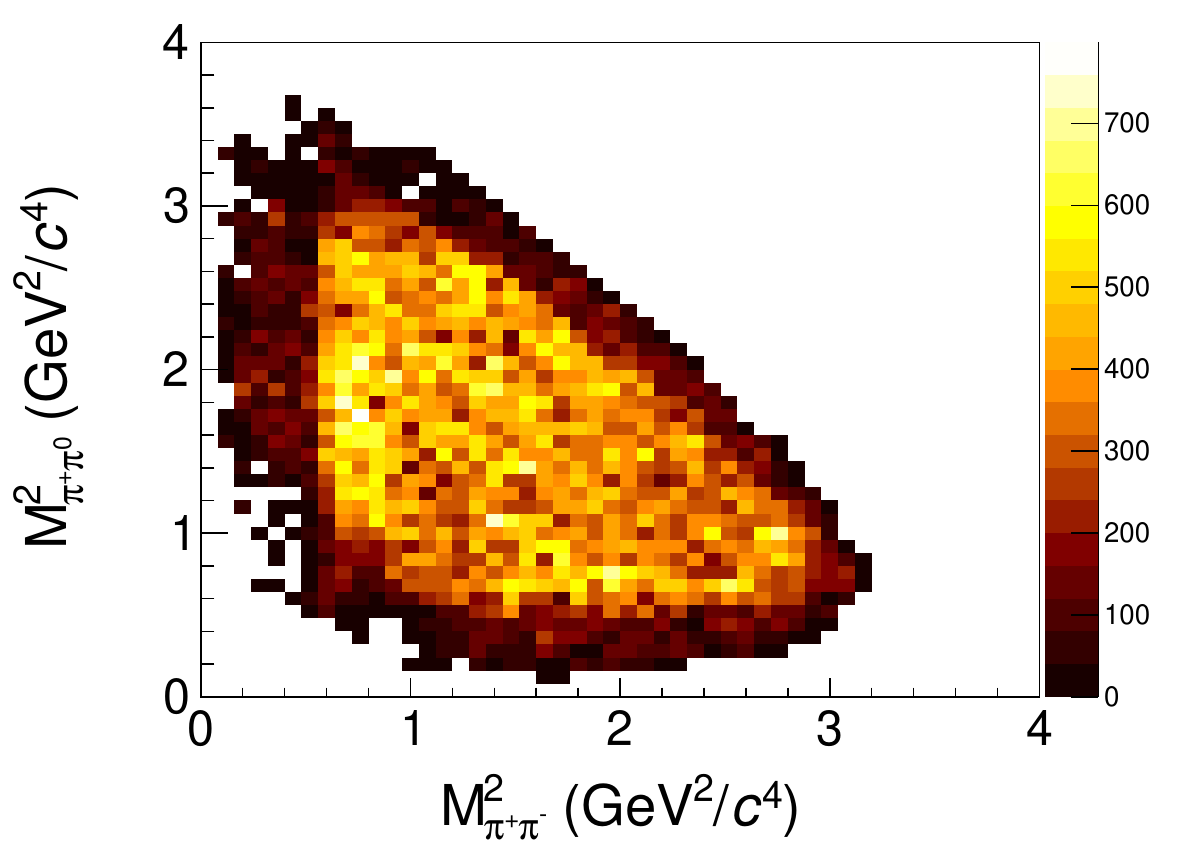}
            \put(60,60){\small\bfseries{(b)}}
        \end{overpic}
        \caption{The Dalitz plots of \mbox{$\processppp$} candidates in (a) data and (b) Phokhara MC sample at \mbox{$\energy$}.}
        \label{figure:dalitz}
    \end{center}
\end{figure}
}

\section{Partial wave analysis}
{
Using the GPUPWA framework~\cite{5_1_pwa_theory_GPUPWA}, a PWA is performed on the surviving candidate events to disentangle the intermediate processes in $\processppp$. The construction of the decay amplitude follows the formalism described in Ref.~\cite{5_1_pwa_theory}, where the decay amplitude of the process $J/\psi\ra\pipipi$ is described with isospin conservation explicitly. In this work, the $J/\psi$ produced from $\ee$ annihilation is replaced by the $\gamma^{\star}$, and the hadron "scale" parameter $Q_{0}$, which is used in the Blatt-Weisskopf barrier factor~\cite{5_1_pwa_theory,5_1_pwa_theory_barrier1,5_1_pwa_theory_barrier2}, is set to be $\sqrt{0.22/3}~{\rm GeV}/c$. In the PWA, the intermediate process through $\rho$, the G parity breaking process through $\omega$, and processes through their excited states are considered according to $J^{PC}$ conservation. The Gounaris-Sakurai model is used for the $\rho$ propagator, which takes the same form as Ref.~\cite{5_1_pwa_theory_GS}. For the $\rho$ excited states, the $\omega$ meson and its excited states, the propagators are taken as energy-dependent Breit-Wigner functions defined by
\begin{linenomath*}
    \begin{flalign}
        BW(\sqrt{s}) =\frac{1}{\sqrt{s}^{2}-M_{R}^{2}+i\sqrt{s}\varGamma_R(\sqrt{s})},
    \end{flalign}
\end{linenomath*}
where $\varGamma_R(\sqrt{s})$ is the energy-dependent width parameterized by
\begin{linenomath*}
    \begin{flalign}
        \varGamma_R(\sqrt{s})=\varGamma_{R}^{0}\cdot\frac{M_{R}^{2}}{\sqrt{s}^2}\cdot(\frac{\varPhi(\sqrt{s})}{\varPhi(M_{R})})^{2l+1},
    \end{flalign}
\end{linenomath*}
where $M_{R}$ and $\varGamma^{0}_{R}$ are the mass and width of the resonances. The phase space factor $\varPhi(\sqrt{s})$ is taken as
\begin{linenomath*}
    \begin{flalign}
        \varPhi(\sqrt{s})=\frac{\sqrt{s}^2+m_{1}^{2}+m_{2}^{2}-2\sqrt{s}m_1-2\sqrt{s}m_1-2m_1m_2}{2\sqrt{s}},
    \end{flalign}
\end{linenomath*}
where $m_1$ and $m_2$ are the mass of daughter particles.

By using the above formalism, the amplitude of the process \mbox{$\processppp$} can be determined as a function of the four-vector momenta of the final state particles. To satisfy this formalism, an one-constraint (1C) kinematic fit and a coordinate transformation are performed for selected signal events in both data and MC sample. The 1C kinematic fit requires the invariant mass of $\gamma\gamma$ equals to the mass of $\piz$ without $\chisq$ requirement. Then, all four-vector momenta of the final state particles are transformed into the rest frame of the $\pip\pim\gamma\gamma$. These modifications are only applied after event selection when four-vector momenta of particles are used as input of PWA model. The complex parameters $\varLambda_{i}$ in Ref.~\cite{5_1_pwa_theory} are free parameters. The mass and width of each resonance are fixed to individual PDG values~\cite{1_PDG}. The magnitude and phase of the complex parameter contribute as two degrees of freedom to each intermediate process. By fixing the magnitude and phase of the process \mbox{$\ee\ra\rho\pi\ra\pipipi$} to 1 and 0, the relative magnitudes and phases of other individual intermediate processes are determined by performing an unbinned maximum likelihood fit using MINUIT~\cite{5_1_pwa_theory_MINUIT}. The likelihood function is
\begin{linenomath*}
    \begin{flalign}
        \mathcal{L}=\prod_{i}^{N}\frac{\omega(\xi_i)\epsilon(\xi_i)}{\sigma_{\rm MC}},
    \end{flalign}
\end{linenomath*}
where $\xi_i$ is the four-vector momenta of the final state particles; $\omega(\xi_i)$ is the differential cross section; $\epsilon(\xi_i)$ is the event selection efficiency, which only depends on four-momenta; the normalization factor $\sigma_{\rm MC}$ is calculated from a MC sample generated with phase space model. The background events from the sideband of $\gamma\gamma$ invariant mass distribution are taken into the likelihood with a weight of $-0.5$.

The fit procedure starts including $\rho$, $\omega$ and all their excited states listed by PDG. The statistical significance of each amplitude is evaluated by incorporating the changes
in likelihood and number of degrees of freedom with and without the corresponding
amplitude being included in the fit. The components with statistical significance less than $5\sigma$ are dropped and the fit is repeated until all the remaining components are statistically significant. All previously removed processes are then reintroduced individually to confirm their non-significance.

The above strategy is performed individually on the data samples at $\energy[2.1250,~2.3960,~{\rm and}~2.9000]$, where the statistics is large enough (Table~\ref{table:background_level}). The components with statistical significance larger than $5\sigma$, i.e. the processes $\ee\ra\rhopi$, $\omega\pi$, $\rhostarpi$, $\rho(1700)\pi$, and $\rho_3(1690)\pi$, are the same for the three data samples. For the other sixteen data samples, the statistical significance of the processes would be smaller due to limited statistics. Therefore, these five processes are included with their relative magnitudes and phases determined by the fit at each energy point independently. Table~\ref{table:significance} shows the results of statistical significances of processes with referenceable mass and width on the PDG. Other processes with small significance are cross-checked by scanning the mass and width values of the intermediate resonances. None of them is found to be significant enough to be included in the PWA model. The fit results at each energy point, including the relative magnitudes and phases of each intermediate process, are summarized in Table~\ref{table:fitresult}.

\begin{table}[!htbp]
    \begin{center}
        \caption{The statistical significance of the tested processes for three data samples. The first column represents the intermediate process.}
        \label{table:significance}
        \begin{tabular}{c|c|c|c}
            \hline\hline
            $\sqrt{s} ({\rm GeV})$ & 2.1250      & 2.3960       & 2.9000       \\
            \hline
            $\rho\pi$              & $>50\sigma$ & $>50\sigma$  & $27\sigma$   \\
            $\rho(1450)\pi$        & $18\sigma$  & $10\sigma$   & $13\sigma$   \\
            $\rho(1700)\pi$        & $6.5\sigma$ & $6.4\sigma$  & $6.6\sigma$  \\
            $\rho_{3}(1690)\pi$    & $14\sigma$  & $9.2\sigma$  & $6.8\sigma$  \\
            $\rho(1570)\pi$        & $1.3\sigma$ & $1.1\sigma$  & $0.13\sigma$ \\
            $\omega\pi$            & $31\sigma$  & $11\sigma$   & $9.7\sigma$  \\
            $\omega(1420)\pi$      & $2.8\sigma$ & $0.21\sigma$ & $1.8\sigma$  \\
            $\omega(1650)\pi$      & $3.5\sigma$ & $0.29\sigma$ & $0.53\sigma$ \\
            $\omega_{3}(1670)\pi$  & $3.4\sigma$ & $1.6\sigma$  & $0.56\sigma$ \\
            \hline\hline
        \end{tabular}
    \end{center}
\end{table}

\begin{table*}[!htbp]
    \begin{center}
        \caption{Relative magnitudes and phases for each intermediate process at each energy point from the PWA fit. The magnitude and phase of the amplitude for the process $\ee\ra\rho\pi$ is fixed to 1 and 0, respectively. The uncertainties are statistical.}
        \label{table:fitresult}
        \begin{tabular}{c|c|c|c|c|c|c|c|c}
            \hline\hline
                             & \multicolumn{4}{c|}{Relative magnitudes} & \multicolumn{4}{c}{Relative phases (rad)}                                                                                                               \\
            \hline
            $\sqrt{s}$ (GeV) & $\rho(1450)\pi$                          & $\omega\pi$                               & $\rho(1700)\pi$ & $\rho_3(1690)\pi$ & $\rho(1450)\pi$ & $\omega\pi$   & $\rho(1700)\pi$ & $\rho_3(1690)\pi$ \\
            \hline
            2.0000           & $1.29\pm0.10$                            & $0.10\pm0.02$                             & $0.86\pm0.09$   & $0.007\pm0.0003$  & $2.59\pm0.05$   & $1.35\pm0.28$ & $-0.53\pm0.04$  & $-2.91\pm0.04$    \\
            2.0500           & $0.91\pm0.40$                            & $0.10\pm0.02$                             & $0.61\pm0.35$   & $0.009\pm0.0031$  & $-3.02\pm0.36$  & $1.50\pm0.34$ & $-1.23\pm0.86$  & $2.71\pm0.30$     \\
            2.1000           & $0.72\pm0.27$                            & $0.12\pm0.02$                             & $0.25\pm0.55$   & $0.004\pm0.0037$  & $-2.98\pm0.82$  & $1.92\pm0.17$ & $-0.04\pm2.30$  & $-1.48\pm0.52$    \\
            2.1250           & $0.76\pm0.05$                            & $0.10\pm0.02$                             & $0.35\pm0.02$   & $0.004\pm0.0005$  & $-3.12\pm0.14$  & $1.82\pm0.16$ & $-0.31\pm0.54$  & $-1.67\pm0.24$    \\
            2.1500           & $1.14\pm0.23$                            & $0.09\pm0.03$                             & $0.35\pm0.32$   & $0.003\pm0.0015$  & $-2.90\pm0.32$  & $2.66\pm0.56$ & $0.42\pm0.76$   & $-1.70\pm0.51$    \\
            2.1750           & $0.99\pm0.11$                            & $0.08\pm0.02$                             & $0.17\pm0.06$   & $0.002\pm0.0002$  & $-2.76\pm0.07$  & $1.75\pm0.39$ & $0.73\pm0.58$   & $-2.02\pm0.18$    \\
            2.2000           & $1.01\pm0.12$                            & $0.11\pm0.01$                             & $0.42\pm0.20$   & $0.003\pm0.0006$  & $-3.02\pm0.21$  & $1.75\pm0.15$ & $0.33\pm0.28$   & $-1.66\pm0.27$    \\
            2.2324           & $0.95\pm0.04$                            & $0.08\pm0.00$                             & $0.57\pm0.04$   & $0.003\pm0.0001$  & $2.98\pm0.03$   & $1.71\pm0.42$ & $-0.09\pm0.35$  & $-1.72\pm0.10$    \\
            2.3094           & $0.92\pm0.07$                            & $0.08\pm0.03$                             & $0.39\pm0.11$   & $0.002\pm0.0005$  & $2.94\pm0.09$   & $1.74\pm0.35$ & $-0.80\pm0.36$  & $-1.69\pm0.10$    \\
            2.3864           & $0.75\pm0.12$                            & $0.07\pm0.01$                             & $0.58\pm0.18$   & $0.002\pm0.0004$  & $2.46\pm0.35$   & $1.46\pm0.22$ & $-0.46\pm0.19$  & $-1.53\pm0.20$    \\
            2.3960           & $0.99\pm0.06$                            & $0.07\pm0.01$                             & $0.51\pm0.09$   & $0.002\pm0.0002$  & $2.92\pm0.13$   & $1.61\pm0.12$ & $-0.07\pm0.13$  & $-1.26\pm0.15$    \\
            2.6444           & $1.12\pm0.13$                            & $0.08\pm0.01$                             & $0.45\pm0.14$   & $0.001\pm0.0002$  & $2.89\pm0.23$   & $1.55\pm0.19$ & $0.28\pm0.27$   & $-0.33\pm0.44$    \\
            2.6464           & $0.93\pm0.13$                            & $0.09\pm0.01$                             & $0.29\pm0.12$   & $0.001\pm0.0002$  & $2.86\pm0.27$   & $2.05\pm0.19$ & $0.37\pm0.39$   & $-0.21\pm0.24$    \\
            2.9000           & $1.07\pm0.10$                            & $0.10\pm0.01$                             & $0.51\pm0.11$   & $0.001\pm0.0001$  & $2.80\pm0.26$   & $1.60\pm0.14$ & $0.34\pm0.30$   & $-0.91\pm0.19$    \\
            2.9500           & $1.34\pm0.29$                            & $0.15\pm0.03$                             & $0.48\pm0.21$   & $0.001\pm0.0004$  & $-2.97\pm0.37$  & $2.05\pm0.39$ & $0.91\pm0.68$   & $-0.94\pm0.33$    \\
            2.9810           & $0.82\pm0.23$                            & $0.10\pm0.03$                             & $0.67\pm0.21$   & $0.001\pm0.0003$  & $2.86\pm0.48$   & $1.04\pm0.52$ & $0.69\pm0.50$   & $-1.28\pm0.44$    \\
            3.0000           & $0.89\pm0.25$                            & $0.18\pm0.03$                             & $0.47\pm0.24$   & $0.001\pm0.0003$  & $2.60\pm0.64$   & $2.94\pm0.20$ & $0.43\pm0.73$   & $-0.84\pm0.67$    \\
            3.0200           & $1.47\pm0.56$                            & $0.13\pm0.04$                             & $0.98\pm0.63$   & $0.002\pm0.0005$  & $2.58\pm0.84$   & $1.77\pm0.43$ & $0.23\pm0.87$   & $-1.18\pm0.59$    \\
            3.0800           & $1.75\pm0.15$                            & $0.13\pm0.02$                             & $1.00\pm0.07$   & $0.001\pm0.0003$  & $2.32\pm0.26$   & $1.70\pm0.22$ & $0.35\pm0.22$   & $-0.19\pm0.22$    \\
            \hline\hline
        \end{tabular}
    \end{center}
\end{table*}

The $\pip\pim$ invariant mass distribution together with the PWA fit projection is shown in Fig.~\ref{figure:pwaresult}. The contributions from $\rho$, $\omega$, and $\rho(1450)$ processes are the most significant, while the $\rho(1700)$ and $\rho_3(1690)$ processes are less significant and contribute marginally to the signal yield.

\begin{figure}[hbpt]
    \begin{center}
        \begin{overpic}[width=0.45\textwidth]{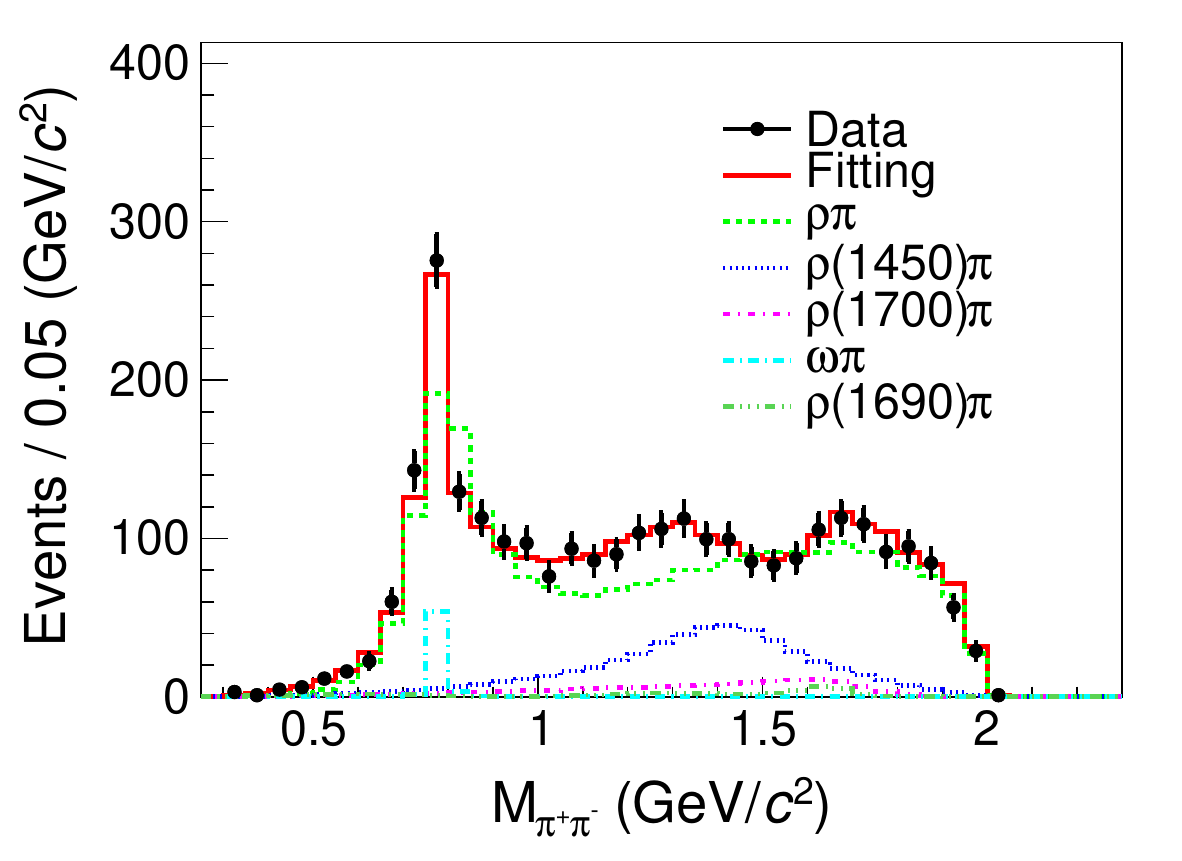}
        \end{overpic}
        \caption{Invariant mass distribution of $\pip\pim$ at \mbox{$\energy[2.3960]$}. The black dots with error bars represent data. The red solid line represents the PWA fit projection. The colored lines with different dashed styles represent the contributions from different intermediate processes.}
        \label{figure:pwaresult}
    \end{center}
\end{figure}

}

\section{Born cross section measurement}
 {
  The $\processppp$ signal yield is obtained by fitting the $\gamma\gamma$ invariant mass distribution with an unbinned maximum likelihood method. The background is described by a first-order polynomial function, and the $\piz$ signal is described by the MC-simulated shape convolved with a Gaussian function, which accounts for the resolution difference between data and MC simulation. The fit result at $\energy$ is shown in Fig.~\ref{figure:signalyield}.

  The signal yields of the $\ee\ra\rhopi\ra\pipipi$ and $\ee\ra\rhostarpi\ra\pipipi$ channels are calculated by
  \begin{linenomath*}
      \begin{flalign}
          N_{\ee\ra\rhopi}=N_{\ee\ra\pipipi}\cdot\mathcal{F}_{\rhopi},
      \end{flalign}
  \end{linenomath*}
  and
  \begin{linenomath*}
      \begin{flalign}
          N_{\ee\ra\rhostarpi}=N_{\ee\ra\pipipi}\cdot\mathcal{F}_{\rhostarpi},
      \end{flalign}
  \end{linenomath*}
  where $N_{\ee\ra X}$ represents the signal yield of the process $\ee\ra X$ and the fractions of the intermediate processes $\mathcal{F}$ are taken from the PWA. The signal yields incorporate the contributions from charged and neutral channels, and the interference between them.

  The detection efficiencies for the $\processppp$ process are estimated using the signal MC samples weighted to the PWA results. The signal MC samples are also weighted according to the differential cross section for intermediate processes $\ee\ra\rhopi$ and $\rhostarpi$ to get their detection efficiencies.

  The Born cross sections of $\processppp$ are calculated by
  \begin{linenomath*}
      \begin{flalign}
          \label{equa:borncrosssection_ppp}
           & \born(\processppp)=\frac{N_{\rm signal}}{\mathcal{L}\cdot\varepsilon\cdot\mathcal{B}_{\piz\ra\gamma\gamma}\cdot (1+\delta)},
      \end{flalign}
  \end{linenomath*}
  where $\mathcal{L}$, $N_{\rm signal}$, $\varepsilon$, and $\mathcal{B}_{\piz\ra\gamma\gamma}$ are the integrated luminosity, signal yield, detection efficiency, and the branching fraction of $\piz\ra\gamma\gamma$ taken from the PDG~\cite{1_PDG}, respectively. The product of the ISR correction factor times the VP correction factor is represented by $(1+\delta)$. The branching fraction $\mathcal{B}_{\rho\rightarrow\pi\pi}=1$ is quoted from the PDG to calculate the Born cross sections of the process $\processrho$. However, the $\mathcal{B}_{\rho(1450)\rightarrow\pi\pi}$ is not available, therefore the Born cross sections of the decay chain $\processrhostar$ are measured.

  The ISR and VP effects are incorporated by ConExc~\cite{4_2_gen_conexc}, depending on the input cross section line shape. An iterative procedure is performed, with comparison between the input cross sections and the measured ones, until the difference of $(1+\delta)$ between the last two iterations is less than 1\%. This method is performed independently for each process. The measured Born cross sections are summarized in Table~\ref{table:borncrosssection}.
  \begin{table*}[!htbp]
      \begin{center}
          \caption{The measured Born cross sections of the process $\processppp$ and of its intermediate processes $\processrho$ and $\processrhostar$. The first uncertainties are statistical and the second ones are systematic.}
          \label{table:borncrosssection}
          \begin{tabular}{c|rPcPl|rPcPl|rPcPl}
              \hline\hline
              $\sqrt{s}$ (GeV) & \multicolumn{3}{c|}{$\born(\processppp)$~(pb)} & \multicolumn{3}{c|}{$\born(\processrho)$~(pb)} & \multicolumn{3}{c}{$\born(\processrhostar)$~(pb)}                                             \\
              \hline
              2.0000           & 425.5                                          & 14.1                                           & 17.7                                              & 419.3 & 18.4  & 16.2 & 107.1 & 13.8 & 6.5 \\
              2.0500           & 358.9                                          & 24.2                                           & 14.5                                              & 459.7 & 72.2  & 22.9 & 59.6  & 39.2 & 3.1 \\
              2.1000           & 363.8                                          & 17.0                                           & 15.7                                              & 403.8 & 119.0 & 17.7 & 32.9  & 16.1 & 2.1 \\
              2.1250           & 398.1                                          & 4.2                                            & 15.6                                              & 452.7 & 7.2   & 24.0 & 43.4  & 5.7  & 2.8 \\
              2.1500           & 379.8                                          & 25.2                                           & 15.3                                              & 447.2 & 70.5  & 19.7 & 99.4  & 37.2 & 5.0 \\
              2.1750           & 367.4                                          & 12.5                                           & 16.2                                              & 423.1 & 21.5  & 18.8 & 71.6  & 17.6 & 3.8 \\
              2.2000           & 331.0                                          & 10.7                                           & 14.5                                              & 345.8 & 32.2  & 15.7 & 60.7  & 12.8 & 3.9 \\
              2.2324           & 311.2                                          & 11.0                                           & 13.4                                              & 320.6 & 13.9  & 14.9 & 50.9  & 3.4  & 3.2 \\
              2.3094           & 243.8                                          & 7.3                                            & 9.6                                               & 250.9 & 27.2  & 12.3 & 38.7  & 1.5  & 2.1 \\
              2.3864           & 181.2                                          & 6.0                                            & 7.7                                               & 167.9 & 15.0  & 7.4  & 17.7  & 4.9  & 1.1 \\
              2.3960           & 179.6                                          & 3.5                                            & 7.7                                               & 165.7 & 7.4   & 7.7  & 31.5  & 3.9  & 1.5 \\
              2.6444           & 93.2                                           & 3.5                                            & 3.8                                               & 75.2  & 5.7   & 3.4  & 19.7  & 4.2  & 1.2 \\
              2.6464           & 91.1                                           & 3.4                                            & 3.6                                               & 74.5  & 5.9   & 3.8  & 13.5  & 3.4  & 0.8 \\
              2.9000           & 50.6                                           & 1.5                                            & 2.4                                               & 36.8  & 2.2   & 2.0  & 9.5   & 1.8  & 0.6 \\
              2.9500           & 39.5                                           & 3.3                                            & 1.7                                               & 30.1  & 4.1   & 1.3  & 12.6  & 4.7  & 0.8 \\
              2.9810           & 49.5                                           & 3.6                                            & 2.0                                               & 33.6  & 4.2   & 1.4  & 5.4   & 2.2  & 0.3 \\
              3.0000           & 42.6                                           & 3.4                                            & 1.7                                               & 31.8  & 4.2   & 1.4  & 5.9   & 2.5  & 0.3 \\
              3.0200           & 34.8                                           & 2.9                                            & 1.5                                               & 19.0  & 3.8   & 0.9  & 9.7   & 5.4  & 0.6 \\
              3.0800           & 20.8                                           & 0.9                                            & 0.9                                               & 8.9   & 0.7   & 0.4  & 6.6   & 1.0  & 0.4 \\
              \hline\hline
          \end{tabular}
      \end{center}
  \end{table*}
 }

\section{Systematic uncertainties}
 {
  Several sources of uncertainties arise from the PWA. These include statistical uncertainties from the fit and systematic uncertainties related to the PWA model. They affect the result by influencing the fractions and the efficiencies of the intermediate processes. The statistical uncertainties are estimated by the standard deviation in percentage of a thousand re-calculated fractions and efficiencies derived from a thousand groups of randomly generated fit parameters using correlated multi-variable Gaussian functions. The parameters and their covariance matrix are taken from the PWA result. The systematic uncertainty related to masses and widths of resonances is estimated by shifting one of them at a time by one standard deviation and performing the fit again. The largest changes caused by all these modifications are taken as the systematic uncertainties. Similarly, the background level is changed by $\pm 50\%$ to estimate the related uncertainty. The energy dependent Breit-Wigner functions and the Gounaris-Sakurai function are replaced by energy independent Breit-Wigner functions to estimate the uncertainty related to the propagator shape. The hadron "scale" parameter $Q_{0}$, which is used in the Blatt-Weisskopf barrier factor, is varied between $0.2$ and $0.33~\unitpgev$ to estimate the associated uncertainty.

  Aside of the PWA, additional sources of systematic uncertainties are considered. The uncertainty associated to the integrated luminosity is 1.0\%~\cite{4_3_error_lumin}. The detection efficiency uncertainty is 1.0\% for each charged track~\cite{3_3_kk,4_3_error_photon} and photon~\cite{4_3_error_photon}. The uncertainty associated to the $\piz\ra\gamma\gamma$ branching fraction is taken from the PDG. The uncertainty related to the 4C kinematic fit is estimated by the changes in result if helix parameters of the simulated charged tracks are corrected to match the resolution in data~\cite{4_3_error_helix}. The systematic uncertainties related to the $E/p$ ratio, opening angle, and helicity angle cuts are estimated by using a control sample of $J/\psi\ra\pipipi$ data. The difference between the detection efficiencies of these selections on data and MC are taken as the uncertainties. The uncertainties for the signal yield and for the background shape are estimated by replacing them by a Gaussian function and by a second-order polynomial function, respectively, and estimating the difference in yields. The uncertainty related to the fit range is estimated by changing it by $\pm 10~\unitmmev$. The uncertainty associated to the $(1+\delta)$ factor is obtained from the accuracy of the radiation function, which is about 0.5\%~\cite{4_3_error_isr}, and the contribution from the cross section line shape, which is estimated by resampling the parameters describing it using correlated multi-variable Gaussian functions. The standard deviation of the recalculated results is taken as the uncertainty. Adding in quadrature all the individual contributions yields the total systematic uncertainty, which varies between $3\%$ and $6\%$ at different energies.
 }

\section{Line shape fitting to the cross section}
{
\begin{figure*}[htbp]
    \begin{center}
        \includegraphics[width=0.90\textwidth]{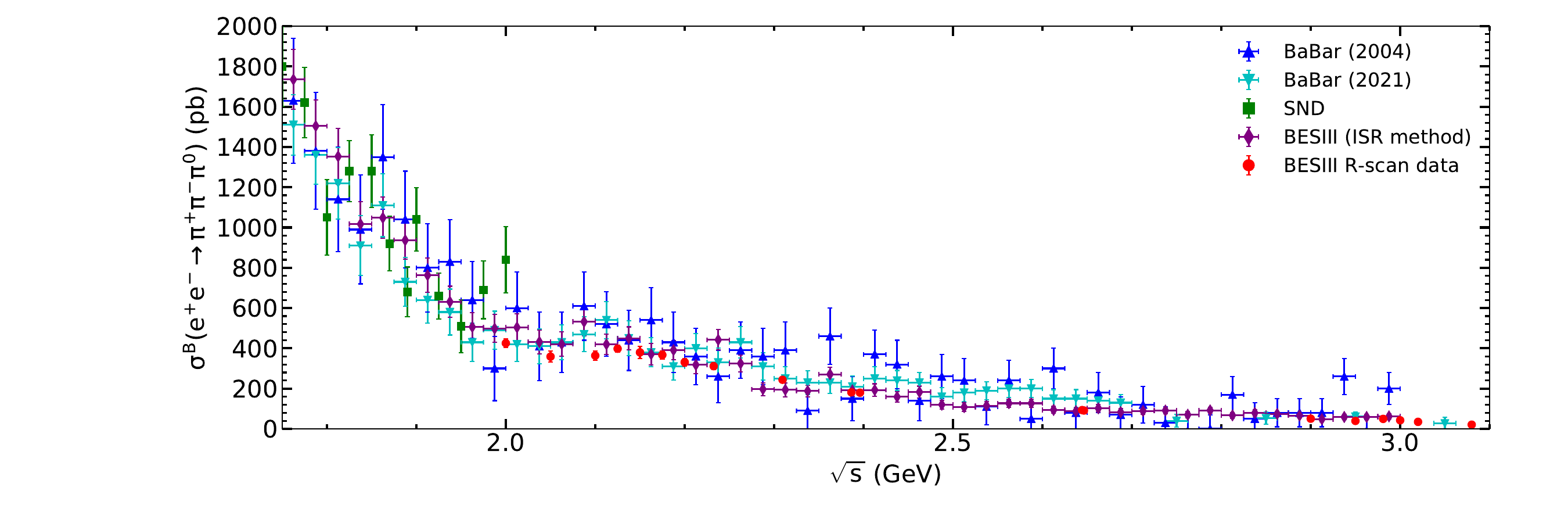}
        \caption{The Born cross-section distribution for the \mbox{$\processppp$} process. The red dots with error bars represent the results obtained in this work. The blue and cyan triangles, green squares, and purple diamonds with error bars represent the \babar, SND and BESIII results measured with the ISR method, respectively. The error bars incorporate both statistical and systematic uncertainties.}
        \label{figure:section:pipipi}
    \end{center}
\end{figure*}
\begin{figure*}[htbp]
    \begin{center}
        \includegraphics[width=0.90\textwidth]{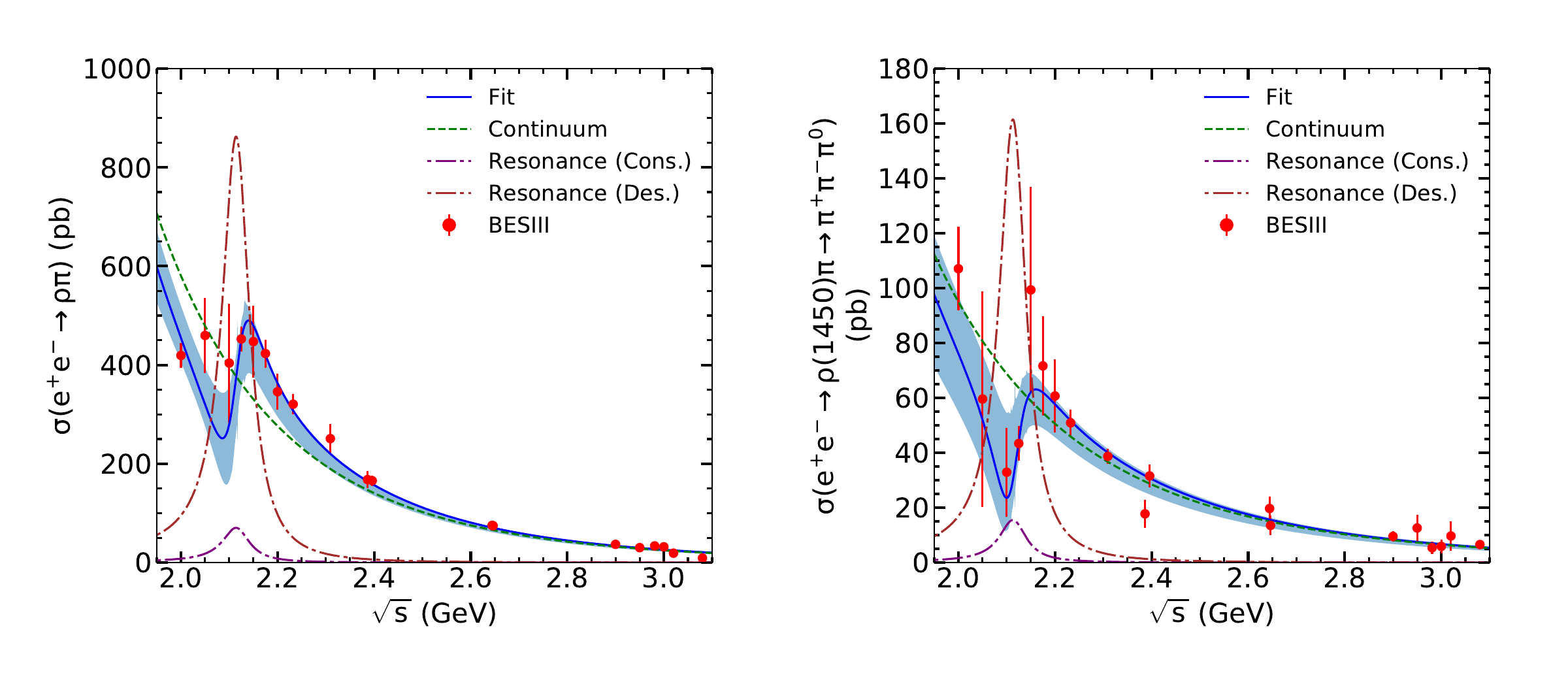}
        \caption{The Born cross-section distributions for the \mbox{$\processrho$} and \mbox{$\processrhostar$} processes. The red dots with error bars represent data. The blue solid, green dashed, purple dot-dashed, and brown dot-dashed lines represent the fit function, the continuum amplitude and the resonance amplitude for the constructive and destructive interference cases, respectively. The blue shaded bands indicate the one-sigma variance of the fit results. The error bars incorporate both statistical and systematic errors.}
        \label{figure:fit:simultaneous}
    \end{center}
\end{figure*}
\begin{figure*}[htbp]
    \begin{center}
        \includegraphics[width=0.90\textwidth]{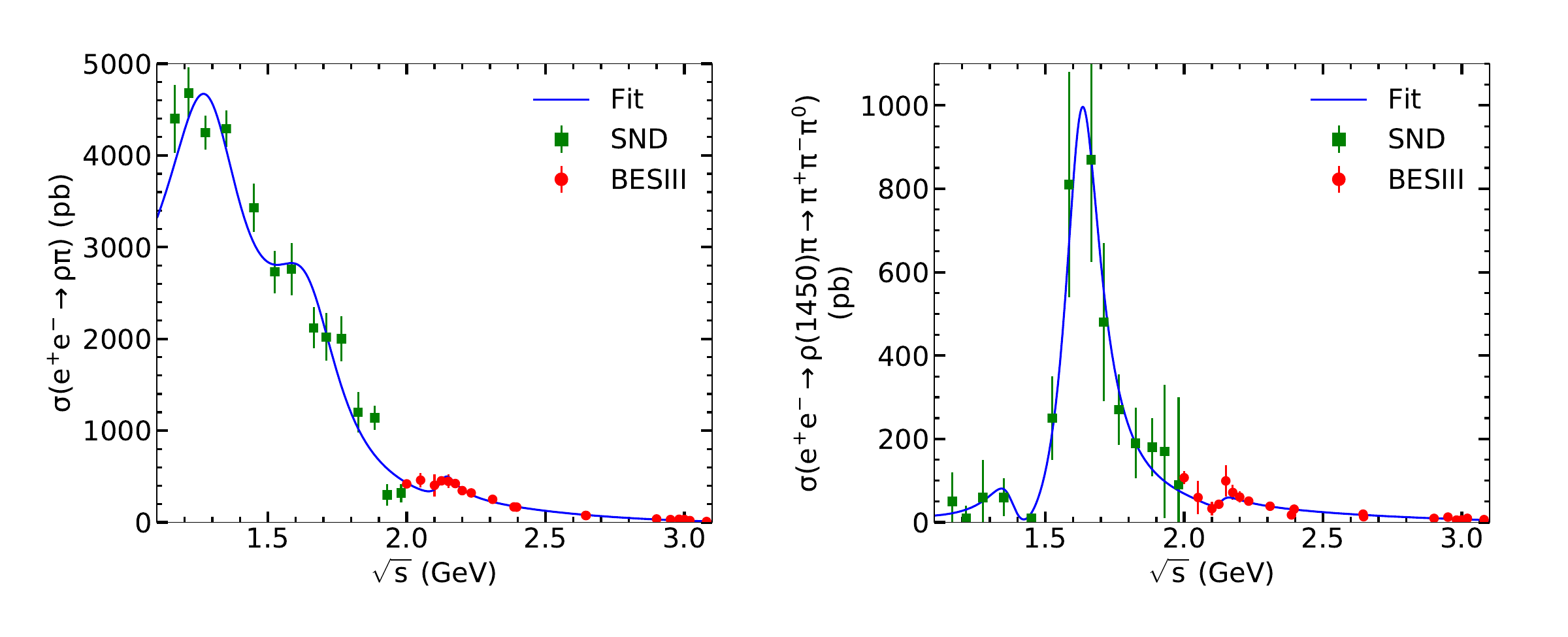}
        \caption{Fit to the \mbox{$\born(\processrho)$} and \mbox{$\born(\processrhostar)$} distributions, merging the BESIII data (red circles with error bars) together with the SND data (green squares with error bars). The blue solid line represent the fit function. The error bars incorporate both statistical and systematic errors.}
        \label{figure:fit:simultaneous_snd}
    \end{center}
\end{figure*}
The measured Born cross sections for the \mbox{$\processppp$} process are shown in Fig.~\ref{figure:section:pipipi}, in agreement with the previous measurements. The Born cross sections for the \mbox{$\processrho$} and \mbox{$\processrhostar$} processes are shown in Fig.~\ref{figure:fit:simultaneous}. The result for the $\ee\ra\omega\piz$ process is not reported, since it has been previously measured by BESIII with better uncertainties by using the same data samples with the subsequent decay \mbox{$\omega\ra\pip\pim\piz$}~\cite{3_3_omegaetaomegapi}. To study the possible structure observed in both of the Born cross section line shapes, a simultaneous fit, assuming the same resonant structures in the the \mbox{$\processrho$} and \mbox{$\processrhostar$} processes, is performed to the measured Born cross sections.
The Born cross-section line shape is described by the contributions from an excited $\omega^{*}$ meson and a continuum function as \mbox{$|f_r(s)e^{i\phi_r}+f_c(s)|^2$}, where $\phi_{r}$ is the relative phase between the resonant and nonresonant amplitudes. The resonance contribution $f_r(s)$~\cite{4_4_lineshape_function} is defined as
\begin{linenomath*}
    \begin{flalign}
        \label{equa:resonance}
        f_r(s)=\frac{M_r}{\sqrt{s}}\frac{\sqrt{12\pi(\hbar c)^{2}\varGamma_{ee}^{r}Br\varGamma_r}}{s-M_{r}^{2}+iM_r\varGamma_r}\sqrt{\frac{\varPhi(\sqrt{s})}{\varPhi (M_r)}},
    \end{flalign}
\end{linenomath*}
where $M_{r}$ and $\varGamma_{r}$ are the mass and width of the structure near $2.20\unitegev$, respectively; $\hbar$ is the reduced Planck's constant; $c$ is the speed of light; $\varPhi$ is the calculated phase space factor. The continuum amplitude $f_c(s)$ is parametrized by $a$ and $b$ as
\begin{linenomath*}
    \begin{flalign}
        f_c(s)=a\frac{\sqrt{\varPhi(\sqrt{s})}}{(\sqrt{s})^{b}}.
    \end{flalign}
\end{linenomath*}

A $\chi^2-$fit, incorporating the correlated and uncorrelated uncertainties among different energy points, is performed. The uncertainties related to luminosity, tracking efficiency, photon detection efficiency, and branching fractions are considered as correlated between different energy points. The nominal fit result is shown in Fig.~\ref{figure:fit:simultaneous}. The significance of the resonance is estimated by comparing the $\chisq$ values of the fit with and without the resonance.

Multiple solutions were found with the same $\chi^2$ depend on the constructive and destructive interferences. These solutions have same mass and width of the resonance structure but different relative phase and electric partial witdh times the branching fraction of the resonance decaying to corresponding final states ($\varGamma_{ee}^{r}Br$). For the \mbox{$\processrho$} process, solutions have $\varGamma_{ee}^{r}Br_{r\ra\rho\pi}=1.5\pm0.7\ {\rm eV}$ and $17\pm12\ {\rm eV}$ for the constructive and destructive interference cases, respectively. For the \mbox{$\processrhostar$} process, they are $\varGamma_{ee}^{r}Br_{r\ra\rhostarpi\ra\pipipi}=0.3\pm0.2\ {\rm eV}$ and $3\pm2\ {\rm eV}$.

To study the systematic uncertainties of the resonance parameters, an alternative fit is carried out by adding the SND measurement~\cite{3_2_snd4} to the BESIII data and the contributions from $\omega(1420)$ and $\omega(1650)$ to our function, which have the same form of $f_r^i(s)e^{i\phi_r^i}$. The alternative fit is shown in Fig.~\ref{figure:fit:simultaneous_snd}. The goodness of the fit is limited by the data samples above $2.2~\unitegev$. The SND data helps to constrain the continuum shape from resonance below $2.0\unitegev$. The difference between the nominal and the alternative fit results is taken as the systematic uncertainty of the resonance parameters. These fit results are summarized in Table~\ref{table:fit:result}.

\begin{table}[!htbp]
    \begin{center}
        \caption{Fit results of the \mbox{$\processrho$} and \mbox{$\processrhostar$} processes, where the uncertainties are statistical only.}
        \label{table:fit:result}
        \begin{small}
            \resizebox{\linewidth}{!}
            {
                \begin{tabular}{c|c|c}
                    \hline\hline
                                              & Nominal fit                                        & Alternative fit      \\
                    \hline
                    Mass                      & $\resultpppmass\pm\resultpppmasserrsta\unitmmev$   & $2134\pm14\unitmmev$ \\
                    Width                     & $\resultpppwidth\pm\resultpppwidtherrsta\unitemev$ & $74\pm31\unitemev$   \\
                    $\chisq$                  & 41.9                                               & 73.3                 \\
                    Number of free parameters & 10                                                 & 18                   \\
                    Degrees of freedom (ndf)  & 28                                                 & 48                   \\
                    $\chisq/{\rm ndf}$        & 1.50                                               & 1.53                 \\
                    Significance              & 5.9$\sigma$                                        & 5.2$\sigma$          \\
                    \hline\hline
                \end{tabular}
            }
        \end{small}
    \end{center}
\end{table}
}

\section{Conclusion}
 {
  In summary, the Born cross sections of \mbox{$\processppp$} are measured, which are consistent with previous measurements within $1\sigma$ with improved precision. A structure with mass \mbox{\resultpppcodemass}\ and width \mbox{\resultpppcodewidth}\ is observed with a significance larger than $5\sigma$ over the continuum-only hypothesis by fitting the Born cross section line shapes for \mbox{$\ee\ra\rho\pi$} and \mbox{$\ee\ra\rho(1450)\pi$}, where the first uncertainties are statistical and the second ones are systematic. According to G-parity conservation in the strong interaction, this resonance decaying to $\rho\pi$ should have a negative G-parity, as an $\omega$ or $\phi$ excited state. Since \mbox{$\phi\ra\pipipi$} is an Okubo-Zweig-Iizuka (OZI) suppressed process, an $\omega$ excited state is a possible candidate. Further experimental studies are needed to clarify the situation. The Born cross sections and PWA results in this work are of importance in the precision determination of the HVP contribution to $a_\mu$ in the near future, as they set the basis for a more accurate description of the \mbox{$\processppp$} process dynamics in MC generators like Phokhara.

  \par
 }

\section{Acknowledgements}
 {
  The BESIII Collaboration thanks the staff of BEPCII and the IHEP computing center for their strong support. This work is supported in part by National Key R\&D Program of China under Contracts Nos. 2020YFA0406400, 2020YFA0406300; National Natural Science Foundation of China (NSFC) under Contracts Nos. 11635010, 11735014, 11835012, 11935015, 11935016, 11935018, 11961141012, 12025502, 12035009, 12035013, 12061131003, 12192260, 12192261, 12192262, 12192263, 12192264, 12192265, 12221005, 12225509, 12235017; the Chinese Academy of Sciences (CAS) Large-Scale Scientific Facility Program; the CAS Center for Excellence in Particle Physics (CCEPP); Joint Large-Scale Scientific Facility Funds of the NSFC and CAS under Contract No. U1832207; CAS Key Research Program of Frontier Sciences under Contracts Nos. QYZDJ-SSW-SLH003, QYZDJ-SSW-SLH040; 100 Talents Program of CAS; The Institute of Nuclear and Particle Physics (INPAC) and Shanghai Key Laboratory for Particle Physics and Cosmology; European Union's Horizon 2020 research and innovation programme under Marie Sklodowska-Curie grant agreement under Contract No. 894790; German Research Foundation DFG under Contracts Nos. 455635585, Collaborative Research Center CRC 1044, FOR5327, GRK 2149; Istituto Nazionale di Fisica Nucleare, Italy; Ministry of Development of Turkey under Contract No. DPT2006K-120470; National Research Foundation of Korea under Contract No. NRF-2022R1A2C1092335; National Science and Technology fund of Mongolia; National Science Research and Innovation Fund (NSRF) via the Program Management Unit for Human Resources \& Institutional Development, Research and Innovation of Thailand under Contract No. B16F640076; Polish National Science Centre under Contract No. 2019/35/O/ST2/02907; The Swedish Research Council; U. S. Department of Energy under Contract No. DE-FG02-05ER41374.
  This paper is also supported by the NSFC under Contracts Nos. 11335008, 11625523, 11705192, 11950410506, 12105276, 12122509; the joint Large-Scale Scientific Facility Funds of the NSFC and CAS under Contracts Nos. U1732263, U1832103, U2032111.
 } \bibliographystyle{apsrev4-1}
\bibliography{main.bib}
\newpage
\end{document}